\documentclass[amsmath,amssymb,aps,pra,twocolumn,longbibliography]{revtex4-1}
\PassOptionsToPackage{breaklinks}{hyperref}
\PassOptionsToPackage{breaklinks}{hyperref}
\AtBeginDocument{\RequirePackage{hyperref}}
\usepackage{tikz}

\usepackage{hyperref}
\usepackage{graphicx}
\usepackage{verbatim}
\usepackage[utf8]{inputenc}

\newcommand{\vp}{\varphi}
\DeclareMathOperator{\tr}{tr}
\newcommand{\vt}{\vartheta}
\newcommand{\tl}{\tilde}

\newcommand{\dl}{\delta}
\DeclareMathOperator{\sinc}{sinc}
\begin{document}
\title{Wigner functions for the pair angle and orbital angular momentum: \\
Possible applications in quantum information}
\author{H. A. Kastrup}
\email{hans.kastrup@desy.de}
\affiliation{DESY Hamburg, Theory Group, Notkestrasse 85, D-22607 Hamburg, Germany}
\begin{abstract}  The framework of Wigner functions for  the canonical pair angle and orbital angular momentum,
 derived and analyzed in 2 recent papers [H. A. Kastrup, Phys. Rev. A {\bf 94}, 062113(2016) and Phys. Rev. A {\bf 95}, 052111(2017)],
is applied to elementary concepts of quantum information like qubits and 2-qubits, e.g., entangled EPR/Bell states etc..
Properties of the associated Wigner functions of such superposed states (pure and mixed) are discussed and illustrated. The Wigner functions of EPR/Bell states are distinguished by their topologically
``twisted'' domain on the configuration subspace $\mathbb{S}^1 \times \mathbb{S}^1$, a torus, of the total phase space. Like the applications of Wigner functions in quantum optics the results presented in the present paper may be useful for the description and analysis of
quantum information experiments with orbital angular momenta of light beams or electron beams, respectively.
\end{abstract} \maketitle
\section{Introduction}
In two recent papers \cite{ka2,ka4} basic properties of Wigner functions on
cylindrical phase spaces $\mathbb{S}^1 \times \mathbb{R}$ (angle and orbital angular momentum, denoted by ``A-OAM'' in the following)
were derived and discussed. The possible usefulness of that concept has, of course, to be demonstrated
by its applications to special systems and associated experiments.\\  A few simple typical example were discussed in
Ch.\ IV C of Ref. \cite{ka2}. The present paper suggests possible applications to such elementary
concepts as ``qubits'' and ``2-qubits'' of quantum information, see, e.g., Refs. \cite{chu,barn}.

The quantized canonical system of the pair {\it angle and orbital angular momentum} is of special
theoretical interest for {\it quantum information} because it provides as a basic framework an infinite dimensional Hilbert space
 $L^2(\mathbb{S}^1, d\vp/2\pi;\dl)$, with orthonormal
basis
\begin{equation}
  \label{eq:6}
   e_{m,\dl}(\vp) = e^{i(m+\dl)\vp}, \;\; m \in \mathbb{Z},\; \dl \in [0,1),
\end{equation}
with  scalar product
\begin{equation}
  \label{eq:1}
  (\psi^{[
    \dl]}_2,\psi^{[\dl]}_1) = \int_{-\pi}^{\pi}\frac{d\vp}{2\pi}\psi^{[\dl]\ast}_2(\vp)\psi^{[\dl]}_1(\vp),\;\;(e_{m,\dl},e_{n,\dl}) = \delta_{mn}
\end{equation} ($\dl_{mn}$: Kronecker symbol) and the expansions
\begin{align}
  \label{eq:2}
  \psi^{[\dl]}(\vp) = & \sum_{m\in \mathbb{Z}} c_m\,e_{m,\dl]}(\vp),\;\; c_m = (e_{m,\dl} \psi), \\ &\psi^{[\dl]}(\vp + 2\pi) = e^{i2\pi\dl}\psi^{[\dl]}(\vp). \nonumber
\end{align}
The number $\dl \in [0,1)$ mathematically characterizes a covering of the group $U(1)$ and physically a fractional OAM \cite{ka}. The following discussions
assume $\dl = 0$ and denote  $L^2(\mathbb{S}^1, d\vp/2\pi) \equiv  L^2(\mathbb{S}^1, d\vp/2\pi;\dl=0),\, e_m \equiv e_{m,\dl =0}$ etc..
The essential generalizations of the main results obtained for $\dl =0$ for
the case $\dl \neq 0$ are discussed in Appendix A.
Note that the coefficients $c_m$ in Eq.\ \eqref{eq:2}
are independent of $\dl$.

One of the advantages of A-OAM systems for quantum information theories is that one can select finite dimensional subspaces of any dimension $d$:
$d=2$: qubits, $d=3$: qutrits, $\ldots,~d$: ``qudits'', like, e.g.,
\begin{equation}
   \label{eq:5}
  (e_{m_0} + e_{m_1} + \ldots + e_{m_{d-1}})/\sqrt{d}
\end{equation} and associated  tensor product spaces which then contain entangled states. (For a recent general discussion of entanglement in quantum theory
see Ref.\cite{witt}.) 

In those $d$-dimensional subspaces $\vp$-independent (``global'') unitary transformations $U(d)$ and other linear mappings (called ``gates'' in quantum information) may act.

Especially one can incorporate the usual elementary qubits from 2-dimensional spaces \cite{barn1} , e.g.,
\begin{equation}
  \label{eq:3}
  (|0\rangle \pm |1\rangle)/\sqrt{2},
\end{equation}
and associated entangled EPR/Bell product states \cite{barn1}
\begin{align}
  \label{eq:4}
  (|00\rangle \pm |11\rangle)/\sqrt{2} \equiv & (|0\rangle\otimes |0\rangle \pm |1\rangle \otimes |1\rangle)/\sqrt{2},\\
(|01\rangle \pm |10\rangle)/\sqrt{2} \equiv & (|0\rangle\otimes |1\rangle \pm |1\rangle \otimes |0\rangle)/\sqrt{2} \label{eq:32}
\end{align} etc.

 In the following the A-OAM Wigner functions of the most general qubits and 2-qubits  will be derived and
some special cases discussed and illustrated in more detail. The discussion is mainly restricted to pure states. The generalization
to mixed states is indicated by Eqs.\ \eqref{eq:55} -- \eqref{eq:56} below.
\subsection*{Experiments}
Experimentally  A-OAM systems have been
investigated particularly with (Laguerre - Gauss) laser light beams
(see, e.g., the reviews \cite{all,pad,and,optorb}) and with
electron beams \cite{evort}. For recent related experiments see, e.g., Refs.\ \cite{rego,chen}.

The crucial property of such beams is that they carry OAM $\vec{p}$ along their directions, i.e., those beams ``rotate'' around their ``axis''!
For experimental investigations of {\em entangled} OAM states see,
e.g., the articles  \cite{zeil1,van,zeil2} and references therein.

The use of associated A-OAM Wigner functions may be helpful for descriptions and analyses of those experiments!
Recall that, in principle, {\it all}  statistical properties of a {\it quantum state of a system can be derived from its Wigner function on the associated
 classical phase space}! In possible applications one should integrate over the radial coordinate of the cylindrical system \cite{ka2} before employing the
A-OAM Wigner function framework.

\section{Wigner functions for qubits}
\subsection{Pure states}
The most general qubit of a A-OAM system is given by
\begin{align}
  \label{eq:7}
&  \chi_{m_0,m_1}^{(\alpha, \beta)}(\vp) = \cos \beta\, e_{m_0}(\vp) + \sin \beta\, e^{i \alpha}\,e_{m_1}(\vp),\\ & m_0,m_1 \in \mathbb{Z},\; m_1 \neq m_0;\, \beta \in (0,\frac{\pi}{2}); \, \alpha \in [0, 2\pi), \nonumber \\
& |\chi_{m_0,m_1}^{(\alpha, \beta)}(\vp)|^2 = 1+ \sin 2\beta \cos[(m_0-m_1)\vp-\alpha]. \label{eq:47}
\end{align}
The states \eqref{eq:7} are elements of the 2-dimensional subspace
\begin{equation}
  \label{eq:27}
  Q^2_{m_0, m_1} = \{\chi_{m_0,m_1} = c_0\,e_{m_0} + c_1\,e_{m_1},\;|c_0|^2+ |c_1|^2=1 \}
\end{equation}
of the overall Hilbert space  $L^2(\mathbb{S}^1, d\vp/2\pi)$.
Here we assume the  correspondences
\begin{equation}
  \label{eq:60}
  |0\rangle \leftrightarrow e_{m_0},\,\, |1\rangle \leftrightarrow e_{m_1}
\end{equation}
for the bases of the 2-dimensional space of qubits and the space $Q^2_{m_0,m_1}$.

The angular momentum operator ($\hbar = 1$ in the following)
\begin{equation}
  \label{eq:61}
 L = (1/i)\partial_{\vp} 
\end{equation}
  has the - obvious - expectation value
\begin{equation}
  \label{eq:8}
  (\chi_{m_0,m_1}^{(\alpha, \beta)},L\chi_{m_0,m_1}^{(\alpha, \beta)}) = m_0\cos^2 \beta + m_1\sin^2\beta,
\end{equation}
which for  $m_1 =- m_0$ becomes 
\begin{equation}
  \label{eq:89}
(\chi_{m_0,-m_0}^{(\alpha, \beta)},L\chi_{m_0,-m_0}^{(\alpha, \beta)})  = m_0\,\cos 2\beta\,.
\end{equation}
This vanishes for $\beta = \pi/4$.

If 
\begin{equation}
  \label{eq:19}
\chi_{m_0,m_1}^{(\hat{\alpha},\hat{\beta})}(\vp)=\cos \hat{\beta}\, e_{m_0}(\vp) + \sin \hat{\beta}\, e^{i\hat{ \alpha}}\,e_{m_1}(\vp)  
\end{equation}
  is another qubit of  the type \eqref{eq:7} in the same 2-dimensional space, then the scalar product of both is given by
\begin{align}
  \label{eq:17}
  &(\chi_{m_0,m_1}^{(\alpha, \beta)},\chi_{m_0,m_1}^{(\hat{\alpha},\hat{\beta})}) =\\& \cos\beta\,\cos\hat{\beta} + e^{-i(\alpha-\hat{\alpha})}\sin\beta \sin\hat{ \beta},
\nonumber \end{align} with the associated transition probability
\begin{align}
  \label{eq:18}
  &|(\chi_{m_0,m_1}^{(\alpha, \beta)},\chi_{m_0,m_1}^{(\hat{\alpha},\hat{\beta})})|^2 =\\& \cos^2\beta\cos^2\hat{\beta} + \sin^2\beta\sin^2\hat{\beta} + \frac{1}{2}\sin2\beta
 \sin2\hat{\beta}\cos(\alpha-\hat{\alpha}), \nonumber
\end{align} which equals $\cos^2(\beta-\hat{\beta)}$ for $\hat{\alpha}= \alpha$. Eq.\ \eqref{eq:18} is of interest in a discussion below (see Eq.\ \eqref{eq:20}).

According to Ch.\ IV of Ref.\ \cite{ka2} the A-OAM Wigner function $V_{\psi}(\theta,p)$ for a wave
function $\psi(\vp)$ from Eq.\ \eqref{eq:2} is given by (with $\hbar$ explicit in the following two Eqs.)
\begin{align}
  \label{eq:9}
  V_{\psi}(\theta,p) & = \frac{1}{2\pi}\int_{-\pi}^{\pi}\frac{d\vt}{2\pi}e^{-i(p/\hbar)\vt}\psi^{\ast}(\theta-\vt/2)\,\psi(\theta + \vt/2) \nonumber \\ &=
\sum_{m,n \in \mathbb{Z}} c_m^{\ast}V_{mn}(\theta,p)c_n\,, \\ V_{mn}(\theta,p) & = \frac{1}{2\pi}e^{i(n-m)\theta}\int_{-\pi}^{\pi}\frac{d\vt}{2\pi}e^{i[(m+n)/2-p/\hbar]\vt}
\label{eq:22} \\ 
& = \frac{1}{2\pi}e^{i(n-m)\theta}\sinc \pi[p/\hbar - (m+n)/2], \nonumber \\ &  \sinc\pi x = \frac{\sin\pi x}{\pi x} = \frac{1}{2\pi}\int_{-\pi}^{\pi}d\vt\,e^{ix\vt}.
\end{align}
 Here $\theta$ describes (by means of the pair $(\cos\theta, \sin \theta)$)
 the points
 on the (configuration) circle $\mathbb{S}^1$ of the classical phase space $\mathcal{P}^2(\theta, p) = \{ (\theta,p) \in \mathbb{S}^1 \times \mathbb{R}\}$, with
 $p \in \mathbb{R}$ the classical canonically conjugate OAM. Integration measure on $\mathcal{P}^2$ is $d\theta dp\; (\hbar =1)$. The letter $"V"$, denoting the Wigner
 function \eqref{eq:9}, stands for ``Vortex''.

The remarkable significance of the sinc-function in the context of the A-OAM Wigner function is elaborately discussed in Ref.\ \cite{ka2}.

Taking for $\psi(\vp)$ the wave function \eqref{eq:7} gives
\begin{align}
  \label{eq:10}
 & 2\pi\,V_{m_0,m_1}^{(\alpha, \beta)}(\theta,p) \\ &=  \cos^2\beta \sinc\pi(p-m_0) + \sin^2\beta \sinc\pi(p-m_1)\nonumber \\
&+ \sin 2\beta \cos[(m_0-m_1)\theta - \alpha]\sinc\pi[p-(m_0 + m_1)/2].\nonumber 
\end{align}

The  last line in Eq.\ \eqref{eq:10} represents the probability interference term in $(\chi_{m_0,m_1}^{(\alpha, \beta)},\chi_{m_0,m_1}^{(\alpha, \beta)})$
-- see Eq.\ \eqref{eq:47} -- and describes, therefore,  essential {\it quantum mechanical} properties of the state!

{\em That means the $\theta$-dependent part of the A-OAM Wig\-ner function
$V_{m_0,m_1}^{(\alpha, \beta)}(\theta,p)$ of the qubit \eqref{eq:7} is solely determined by its probability interference term!}

  That part vanishes
 for  $p - (m_0+m_1)/2 = k,\; k \in \{\mathbb{Z} -\{0\}\}$ and/or 
 if $\alpha$ and $\theta$ are such that $\cos[(m_0-m_1)\theta-\alpha]=0$.
 
On the other hand, the factors of $\cos[(m_0-m_1)\theta - \alpha]$ in Eq. \eqref{eq:10} are maximal $(=1)$ for $\beta = \pi/4$ and $p=(m_0 + m_1)/2$!

The angles $\alpha$ and $\beta$ may depend on other parameters, e.g., time $t$, space coordinates, external fields etc., and may, therefore,
 be manipulated from outside. Their values can be represented
by points on the 2-dimensional surface of a ``Bloch'' sphere \cite{barn2}.
\subsection*{Examples}
It is instructive to look at a few special examples:

1. $m_1 = - m_0$:
\begin{align}
  \label{eq:11}
  & 2\pi\,V_{m_0,-m_0}^{(\alpha, \beta)}(\theta,p)\\ & = \cos^2\beta \sinc\pi(p-m_0) + \sin^2\beta \sinc\pi(p+m_0)\nonumber \\
&+ \sin 2\beta \cos[2m_0\theta - \alpha]\sinc\pi\,p. \nonumber
\end{align}
The special case $\alpha = 0, \beta = \pi/4, m_0 = 1$ of Eq.\ \eqref{eq:11} is  illustrated in
 Fig.\ \ref{fig:fig1c} (see also Fig.\ 2 in Ref.\ \cite{ka2}):
\begin{figure}[h]\includegraphics[width=\columnwidth]{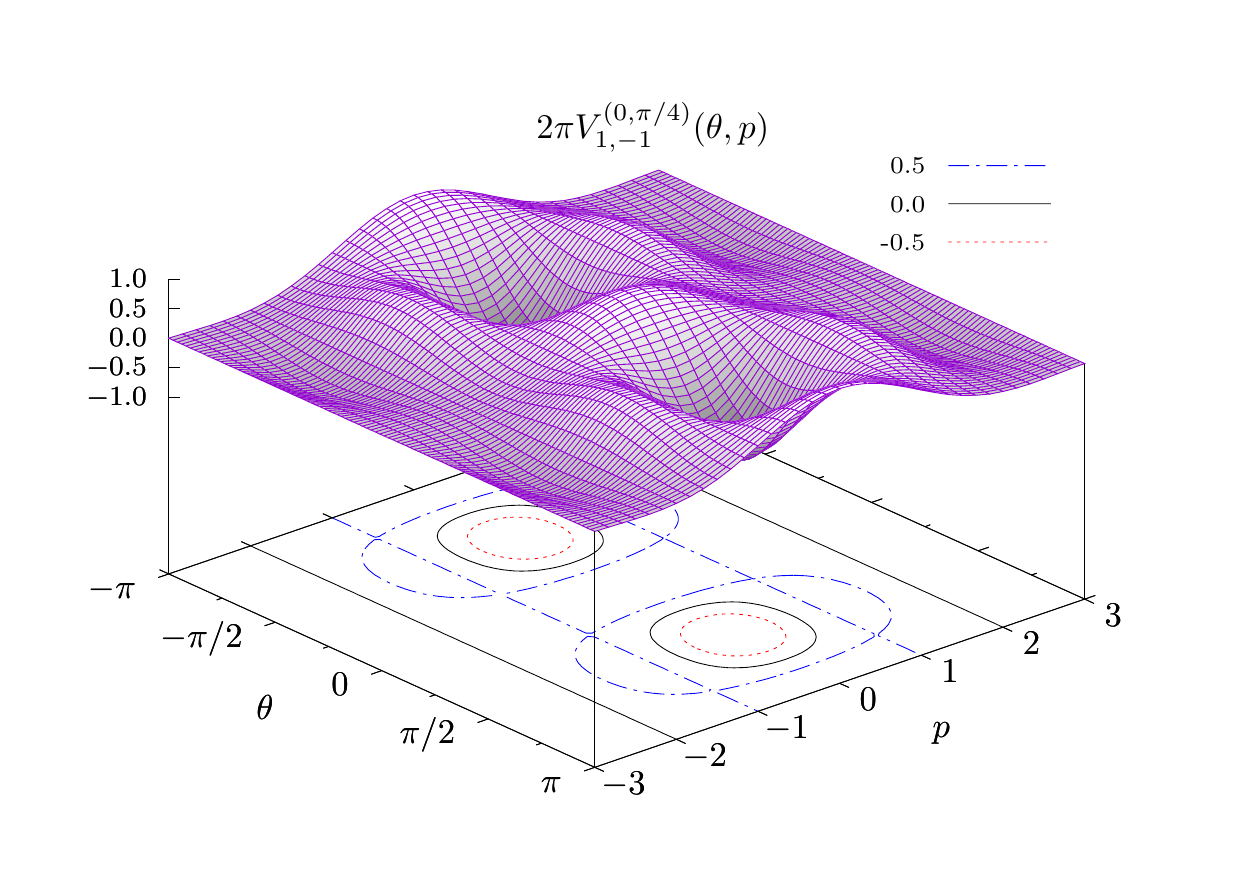}
\caption{\label{fig:fig1c}A-OAM Wigner function $2\pi\, V_{1,-1}^{(\alpha = 0, \beta =\pi/4)} (\theta, p) = \frac{1}{2}[\sinc \pi (p-1) + \sinc\pi (p+1)]
+ \cos 2\theta \, \sinc \pi p$ of the qubit $(e_{+1} + e_{-1})/\sqrt{2}$.} \end{figure}

2. $m_1 = 0$ (ground state of $H \propto L^2$):
\begin{align}
  \label{eq:12}
  & 2\pi\,V_{m_0,0}^{(\alpha, \beta)}(\theta,p)\\ & = \cos^2\beta \sinc\pi(p-m_0) + \sin^2\beta \sinc\pi\,p\nonumber \\
&+ \sin 2\beta \cos[m_0\theta - \alpha]\sinc\pi(p-m_0 /2), \nonumber
\end{align}
Here the interference term vanishes for  $p= m_0/2 + k,\; k \in \{\mathbb{Z}-\{0\}\}$. (Recall that $m_0 \neq 0$ because $m_1 =0$.)

The Wigner function \eqref{eq:12} for the special values $\alpha = 0, \beta = \pi/3, m_0 = 1$ is shown in Fig.\ \ref{fig:fig2c}:
\begin{figure}[h]\includegraphics[width= \columnwidth]{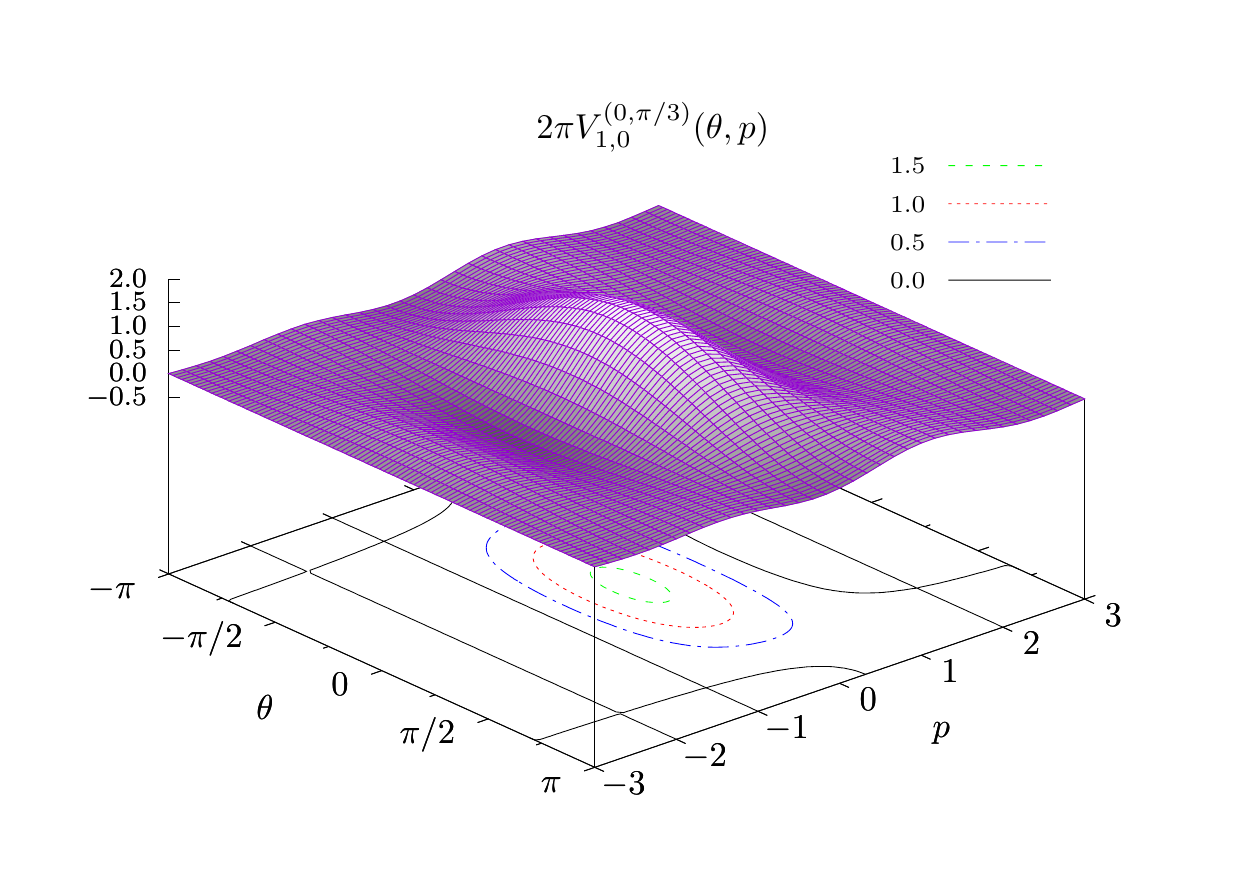}
\caption{\label{fig:fig2c} A-OAM Wigner function $2\pi\, V_{1,0}^{(\alpha=0,\beta = \pi/3)}(\theta,p) = \frac{1}{4}[\sinc\pi(p-1) + 3\sinc\pi p] + \frac{\sqrt{3}}{2}
\cos \theta\,\sinc\pi(p-1/2)$ for the qubit $(e_{+1} +\sqrt{3}\, e_0)/2$.} \end{figure}
\subsection{Marginal probability distributions}
The quantum mechanical {\it marginal} probability distributions $|  \chi_{m_0,m_1}^{(\alpha, \beta)}(\theta)|^2$ (angular distribution density)
and $\{\cos^2\beta,\sin^2\beta\}$ (OAM distribution) of the state \eqref{eq:7} can be obtained, according to Ref.\ \cite{ka2}, from the A-OAM Wigner function
\eqref{eq:10} as follows:
\begin{align}
  \label{eq:13}
 & \int_{-\infty}^{\infty}dp\,V_{m_0,m_1}^{(\alpha, \beta)}(\theta,p)\\ =& \frac{1}{2\pi}\{1+\sin2\beta\,\cos[(m_0-m_1)\theta - \alpha]\} \nonumber \\
= &\frac{1}{2\pi} | \chi_{m_0,m_1}^{(\alpha, \beta)}(\theta)|^2 \nonumber
\end{align} -- compare Eq.\ \eqref{eq:47} --,
where the relation
\begin{equation}
  \label{eq:14}
  \int_{-\infty}^{\infty}dp\,\sinc\pi(p+a) = 1,\; a \in \mathbb{R},
\end{equation}
has been used. Comparing Eqs.\ \eqref{eq:13} and \eqref{eq:47} one has to take into account that in the present case the integration measure for the phase space is $d\theta\, dp$ (with $\hbar =1$), whereas we have $d\vp/(2\pi)$ in the integral \eqref{eq:1} \cite{ka2,ka4}.

Integration over $\theta$ gives Whittaker's cardinal function \cite{ka2}
\begin{align}
  &\label{eq:15}
  \int_{-\pi}^{\pi}d\theta\,V_{m_0,m_1}^{(\alpha, \beta)}(\theta,p) = \omega_{m_0,m_1}^{(\alpha,\beta)}(p)\\&= \cos^2\beta\,\sinc\pi(p-m_0) + \sin^2\beta\,\sinc\pi(p-m_1),
\nonumber \end{align}
from which the quantum mechanical OAM probabilities $\cos^2\beta$ and $\sin^2\beta$ can  be extracted immediately with the help of the orthonormality relations
\cite{ka2}
\begin{equation}
  \label{eq:16}
  \int_{-\infty}^{\infty}dp\,\sinc\pi(p-m)\,\sinc\pi(p-n) = \delta_{mn}.
\end{equation}
If $V_{m_0,m_1}^{(\alpha, \beta)}(\theta,p)$ and $V_{m_0,m_1}^{(\hat{\alpha}, \hat{\beta)}}(\theta,p)$ are A-OAM Wigner functions of
 the qubits \eqref{eq:7} and \eqref{eq:19},
then the transition probability \eqref{eq:18} is now given \cite{ka2} by the integral
\begin{align}
  \label{eq:20}
 &2\pi \int_{-\infty}^{\infty}dp\int_{-\pi}^{\pi}d\theta V_{m_0,m_1}^{(\alpha, \beta)}(\theta,p)V_{m_0,m_1}^{(\hat{\alpha}, \hat{\beta)}}(\theta,p)= \\ &
 \cos^2\beta\cos^2\hat{\beta}\nonumber + \sin^2\beta\sin^2\hat{\beta} + \frac{1}{2}\sin2\beta
 \sin2\hat{\beta}\cos(\alpha-\hat{\alpha}), \nonumber
\end{align}
where the relations \eqref{eq:16}, \eqref{eq:14} and
\begin{equation}
  \label{eq:21}
  \int_{-\infty}^{\infty}dp\,\sinc^2\pi(p+a) = 1,\; a \in \mathbb{R},
\end{equation}
have been used \cite{ka5}.

\subsection{Mixed states}
I briefly indicate the changes if the pure state $\chi_{m_0,m_1}$ of the subspace \eqref{eq:27} is replaced by a mixed state represented by a $2\times 2$ density 
matrix $\rho^{[m_0, m_1]}$ with elements $\rho_{m_j,m_k},\,j,k = 0,1$: Now $V_{\psi}(\theta,p)$ of Eq.\ \eqref{eq:9} is replaced by \cite{ka2,ka4}
\begin{align}
  \label{eq:55}
  V_{\rho^{[m_0,m_1]}}(\theta,p) =& \tr\{\rho^{[m_0,m_1]}\cdot V^{[m_o,m_1]}(\theta, p)\}, \\ & V^{[m_0,m_1]}(\theta, p)= (V_{m_j m_k}(\theta, p)), \nonumber
\end{align}
with the matrix elements $V_{m_j m_k}(\theta,p),\, j,k = 0,1 $ as in Eq.\ \eqref{eq:22}. 

 For a general $2\times 2$ density matrix
\begin{equation}
  \label{eq:57}
  \rho^{[m_0, m_1]} = \frac{1}{2}(I_2 + \vec{a}\cdot\vec{\sigma}],
\end{equation}
with $I_2$  the unit matrix in 2 dimensions, $ \vec{a} \in \mathbb{R}^3, \, \vec{a}^2 \leq 1$ and $\sigma_j,\, j=1,2,3,$ Pauli's matrices, we get
\begin{align}
  \label{eq:58}
 & 2\pi \tr[\rho^{[m_0, m_1]}\cdot V^{[m_0, m_1]}(\theta, p)]\\& =  \frac{1+ a_3}{2} \sinc \pi (p-m_0) + \frac{1-a_3}{2}\sinc \pi(p-m_1) \nonumber \\ & 
+ \{a_1\cos[(m_0-m_1)\theta] +
a_2\sin[(m_0 -m_1)\theta]\}  \nonumber \\ & \times \sinc \pi [p - (m_0 + m_1)/2]. \nonumber
\end{align} Note that here, too, (cf.\ Eq.\ \eqref{eq:10}) the $\theta$-dependent term of $V_{\rho^{[m_0,m_1]}}(\theta,p)$ is solely determined by the interference part of $\rho^{[m_0, m_1]}$
The essential changes of the main results obtained for $\dl =0$ for the case $\dl \neq 0$ are discussed in $\rho^{[m_0, m_1]}$
in Eq.\ \eqref{eq:57}: the numbers $a_1 \pm i\,a_2$ form its anti-diagonal. For the pure case $ \vec{a}^2 =1$ the expression \eqref{eq:58} becomes equivalent to the function \eqref{eq:10}.

 Important for applications is the overlap formula  of two density matrices
\begin{align}
  \label{eq:56}
 & \tr(\rho^{[m_0,m_1]}_{1}\cdot\rho^{[m_0,m_1]}_{2})\\ & = 2\pi\int_{-\infty}^{\infty}d(p/\hbar)\int_{-\pi}^{\pi}d\theta\,{V_{\rho^{[m_0,m_1]}_1}}(\theta,p)\,
{V_{\rho^{[m_0,m_1]}_2}}(\theta,p), \nonumber
\end{align} derived in Refs.\cite{ka2,ka4}. \par
\section{Wigner functions for 2-qubits, especially  EPR/Bell states}
\subsection{The quantum state space of 2-qubits}
For the discussion of the tensor product of the Hilbert space  $L^2(\mathbb{S}^1, d\vp/2\pi)$ from above -- characterized by the Eqs.\
 \eqref{eq:6}--\eqref{eq:2} -- with itself we have to go
slightly beyond the A-OAM framework discussed in Refs. \cite{ka2,ka4}:

 There we had a phase space $\mathcal{P}^2(\theta,p) = 
\{(\theta,p) \in \mathbb{S}^1\times \mathbb{R}\}$ with the circle $\mathbb{S}^1$ as configuration space and the real line $\mathbb{R}$ as cotangent (canonical momentum) space.
 Coordinates for the former are provided by the pair
 $(\cos\theta, \sin\theta)$ and the angular momentum $p \in \mathbb{R}$ for the latter. 
By doubling the system we get the phase space 
\begin{align}
  \label{eq:30}
   \mathcal{P}^4(\tl{\theta}, \tl{p}) =& \{(\theta_1, \theta_2; p_1, p_2) \in \mathbb{S}^1\times \mathbb{S}^1 \times
\mathbb{R}\times \mathbb{R})\}, \\ & \tl{\theta} \equiv \{ \theta_1, \theta_2\}, \;\; \tl{p} \equiv \{p_1,p_2\}. \nonumber
\end{align}
 Configuration space is now the torus $\mathbb{S}^1\times \mathbb{S}^1$.

A crucial tool for the derivation of the Wigner function \eqref{eq:9} in Ref.\ \cite{ka2} are the unitary representations of the Euclidean group $E(2)$ of the
 plane \cite{ka6}.
In our case, $\mathcal{P}^4(\tl{\theta}, \tl{ p})$,  we have to employ the direct product $E(2) \times E(2)$ and the associated unitary representations. The procedure
for deriving the A-OAM Wigner function in question is then stricly analogue to that of Ch.\ II in Ref.\ \cite{ka2} for the expression \eqref{eq:9} above
 and the result is as expected:

We  have the product Hilbert space 
\begin{equation}
  \label{eq:39}
  L^2(\mathbb{S}^1\times \mathbb{S}^1, d\vp_1d\vp_2/(2\pi)^2),
\end{equation}
 with basis
\begin{equation}
  \label{eq:23}
  e_{mn}(\tl{\vp}) = e_m(\vp_1)\,e_n(\vp_2) = e^{im\vp_1 + in\vp_2},\, m,n \in \mathbb{Z},
\end{equation}
scalar product
\begin{align}
  \label{eq:24}
  (\psi_2,\psi_1) =& \int_{-\pi}^{\pi}\frac{d^2\tl{\vp}}{(2\pi)^2}\psi^{\ast}_2(\tl{\vp})\psi_1(\tl{\vp}),\\
&(e_{km},e_{ln}) = \delta_{kl}\,\delta_{mn}, \nonumber
\end{align}
and expansions
\begin{equation}
  \label{eq:25}
  \psi(\tl{\vp}) = \sum_{m,n \in \mathbb{Z}}c_{mn}\,e_{mn}(\tl{\vp}),\,\, c_{mn} = (e_{mn},\psi).
\end{equation}
The functions \eqref{eq:23} are eigenfunctions of the total OAM operator:
\begin{equation}
  \label{eq:38}
  L = \frac{1}{i}\partial_{\vp_1} +\frac{1}{i}\partial_{\vp_2},\;\; Le_{mn} = (m+n)e_{mn}. 
\end{equation}
We here assume for both Hilbert space factors of their tensor product that $\dl_1=0=\dl_2$. See the discussion after Eq.\ \eqref{eq:2}.
The cases $\dl_1, \dl_2 \neq 0$ will be discussed in  Appendix A.

Comparison of the basis \eqref{eq:23}  with the correspondences \eqref{eq:60}  implies the following correspondences for a basis of a 4-dimensional
subspace $Q^4_{m_0m_1,n_0n_1}$ considered as the tensor product of the
space \eqref{eq:27} with itself (without the normalization $|c_1|^2 + |c_2|^2 =1$):
\begin{align}
  \label{eq:33}
 &e_{m_0 n_0}\leftrightarrow |00\rangle,\;  e_{m_1 n_1}\leftrightarrow |11\rangle,\\
&  e_{m_0 n_1}\leftrightarrow |01\rangle,\;  e_{m_1 n_0}\leftrightarrow |10\rangle. \nonumber 
\end{align}
The elements of the 4-dimensional space $Q^4_{m_0m_1,n_0n_1}$, generated by the tensor product of qubits, are called ``2-bits''.
Another possible basis for $Q^4_{m_0m_1,n_0n_1}$ constitute  the four EPR/Bell states \eqref{eq:4} and \eqref{eq:32}.
\subsection{Wigner functions for general 2-qubits}
Applying the same arguments of Ch.\ II in Ref.\ \cite{ka2} -- which lead to the Wigner function \eqref{eq:9} above -- now to the products $\mathcal{P}^4(\tl{\theta},
 \tl{p})$ and $E(2)\times E(2)$  we then get, on the phase space $\mathcal{P}^4$ for the wave function $\psi$ of Eq.\ \eqref{eq:25},
 the A-OAM Wigner function 
\begin{align}
  \label{eq:26}
 & V_{\psi}(\tl{\theta}, \tl{p}) =  \\ & \frac{1}{(2\pi)^2}\int_{-\pi}^{\pi}\frac{d^2\tl{\vt}}{(2\pi)^2}
e^{-i(p_1\vt_1 + p_2\vt_2)}\,\psi^{\ast}(\tl{\theta}-\tl{\vt}/2)\,
\psi(\tl{\theta}+\tl{\vt}/2), \nonumber
\end{align}
which is the obvious generalization of the expression \eqref{eq:9}.
We next determine the A-OAM Wigner functions for general 2-qubit elements
\begin{align}
  \label{eq:28}
  \psi_{2\!-\!qb}(\tl{\vp})& =c_{00}\,e_{m_0 n_0}(\tl{\vp}) +c_{10}\,e_{m_1 n_0}(\tl{\vp})\\  & + c_{01}\,e_{m_0 n_1}(\tl{\vp}) + c_{11}\,e_{m_1n_1}(\tl{\vp}), \nonumber \\
&|c_{00}|^2 + |c_{01}|^2 + |c_{10}|^2 + |c_{11}|^2 = 1. \nonumber
\end{align}
of the 4-dimensional tensor product space 
The four complex coefficients $c_{jk}$ may be parametrized by  real numbers as follows:
\begin{align}
  \label{eq:29}
  &c_{00} = b_{00},\;c_{10}=e^{i\alpha_{10}}b_{10},\;c_{01}=e^{i\alpha_{01}}b_{01},\;c_{11}=e^{i\alpha_{11}}b_{11}, \nonumber \\
& \alpha_{jk}\in [0, 2\pi), \;b_{jk} \in \mathbb{R}, \; b_{00}^2 + b_{10}^2 +b_{01}^2 +b_{11}^2 =1.  
\end{align}
Finally, a convenient parametrization for the real $b_{jk}$ is
\begin{align}
  \label{eq:59}
  &b_{00}= \cos\beta,\; b_{10} = \sin\beta \cos\gamma,\; \beta, \gamma \in [0,\pi) \\ & b_{01} = \sin\beta \sin\gamma \cos\phi,\; b_{11}= \sin\beta \sin\gamma \sin \phi,
  \; \phi \in [0,2\pi).                                                                                      \nonumber
 \end{align} 
 Inserting the wave function \eqref{eq:28} into the expression \eqref{eq:26} yields the most general 2-qubit A-OAM Wigner function {\allowdisplaybreaks
\begin{align}
  \label{eq:31}
  &(2\pi)^2V_{\psi_{2\!-\!qb}}(\tl{\theta},\tl{p}) \\ & = b_{00}^2\,\sinc\pi(p_1-m_0)\sinc\pi(p_2-n_0)\nonumber \\ & +b_{10}^2\sinc\pi(p_1-m_1)\sinc\pi(p_2-n_0) \nonumber\\
&+b_{01}^2\sinc\pi(p_1-m_0)\sinc\pi(p_2-n_1)\nonumber \\ & +b_{11}^2\sinc\pi(p_1-m_1)\sinc\pi(p_2-n_1) \nonumber \\
&+ 2\,b_{00}b_{10}\cos[(m_1-m_0)\theta_1+\alpha_{10}] \nonumber\\
&\times \sinc\pi[p_1-(m_0+m_1)/2]\sinc\pi(p_2-n_0) \nonumber \\
&+ 2\,b_{00}b_{01}\cos[(n_1-n_0)\theta_2+\alpha_{01}] \nonumber\\
&\times \sinc\pi(p_1-m_0)\sinc\pi[p_2-(n_0+n_1)/2] \nonumber \\
&+ 2\,b_{00}b_{11}\cos[(m_1-m_0)\theta_1+ (n_1-n_0)\theta_2 +\alpha_{11}] \nonumber\\
&\times \sinc\pi[p_1-(m_0+m_1)/2]\sinc\pi[p_2-(n_0+n_1)/2] \nonumber\\
&+ 2\,b_{01}b_{10}\cos[(m_1-m_0)\theta_1-(n_1-n_0)\theta_2+\alpha_{10}-\alpha_{01}] \nonumber\\
&\times \sinc\pi[p_1-(m_0+m_1)/2]\sinc\pi[p_2-(n_0+n_1)/2] \nonumber \\
&+ 2\,b_{01}b_{11}\cos[(m_1-m_0)\theta_1+\alpha_{11}-\alpha_{01}] \nonumber\\
&\times \sinc\pi[p_1-(m_0+m_1)/2)]\sinc\pi(p_2-n_1) \nonumber \\
&+ 2\,b_{10}b_{11}\cos[(n_1-n_0)\theta_2+\alpha_{11}-\alpha_{10}] \nonumber\\
&\times \sinc\pi(p_1-m_1)\sinc\pi[p_2-(n_1+n_0)/2]. \nonumber
\end{align}} Essential properties of this general expression can be seen from the discussion of the following special examples:
\subsection{Wigner functions for special 2-qubits}
\subsubsection{Wigner function for the basis vector $e_{m_0n_0}$}
The A-OAM Wigner function for the basis vector $e_{m_0 n_0}(\tl{\vp})$ is given by
\begin{equation}
  \label{eq:35}
  V_{m_0 n_0}(\tl{\theta},\tl{p}) = \frac{1}{(2\pi)^2}[\sinc\pi(p_1-m_0)\sinc\pi(p_2-n_0)].
\end{equation}
This follows from Eq.\ \eqref{eq:31} with $ b_{00}=1, b_{01}=b_{10}=b_{11}=0$.

The Wigner function \eqref{eq:35} is independent of $\theta_1$ and $\theta_2$!

The expression \eqref{eq:35} is exactly the product $V_{m_o}(\theta_1,p_1)\,V_{n_0}(\theta_2, p_2)$ of the Wigner functions for the basis vectors $e_{m_0}$ and $e_{n_0}$
respectively\cite{ka2}.
\subsubsection{Wigner functions of factorizable 2-qubits}
If $c_{00}  = b_{00}= \cos\beta$ and $c_{10} = b_{10}e^{i\alpha_{10}} = \sin\beta e^{i\alpha_{10}} $, $b_{01} =0 = b_{11}$, then we have for the state \eqref{eq:28}
\begin{align}
  \label{eq:62}
 \psi^{(\alpha_{10}, \beta)}_{00,10}(\tl{\vp}) = &  \cos\beta e_{m_0n_0}(\tl{\vp})+ \sin\beta e^{i\alpha_{10}}\,e_{m_1n_0}(\tl{\vp})\\  =&  
[\cos\beta e_{m_0}(\vp_1)+ \sin\beta e^{i\alpha_{10}}\,e_{m_1}(\vp_1)]\,e_{n_0}(\vp_2). \nonumber
\end{align}
According to Eq.\ \eqref{eq:31} the associated Wigner function is given by
\begin{align}
  \label{eq:51}
 & V_{00,10}^{(\alpha_{10}, \beta)}(\tl{\theta},\tl{p})  \\ &= \frac{1}{(2\pi)^2} \{\cos^2\beta \sinc\pi(p_1-m_0) +   \sin^2\beta \sinc\pi(p_1-m_1) \nonumber
 \\ & + \sin 2\beta\cos[(m_1-m_0)\theta_1 +\alpha_{10}]
\nonumber \sinc\pi[p_1 -(m_0+m_1)/2]\}\\ & \times  \sinc\pi(p_2-n_0) = V_{m_0,m_1}^{(\alpha_{10}, \beta)}(\theta_1,p_1)\, V_{n_0}(\theta_2, p_2). \nonumber
\end{align}
Thus, the Wigner function of the product state \eqref{eq:62} is a product of the qubit Wigner function \eqref{eq:10} with $\theta =
\theta_1$  and $p=p_1$  times the Wigner functions $V_{n_0}(\theta_2,p_2) = \sinc\pi(p_2-n_0)/(2\pi)$  of the basis vector $e_{n_0}$  \cite{ka2}.
Corresponding arguments hold for the  special cases $(b_{00}, b_{10}, b_{01}, b_{11})$ = 
 $(\cos\beta, 0, \sin\beta, 0)$, 
 $(0,0,\cos\phi, \sin\phi)$ or $(0,\cos\gamma, 0, \sin\gamma)$. Note that all four cases of the type \eqref{eq:51} do depend on $\theta_1\text{ or } \theta_2$ only!
 \subsubsection{A simplified general case}
 Considerable simplifications are obtained for the terms in expression \eqref{eq:31} if $m_0+m_1 =0,\,n_0+n_1 =0,\; m_0 \neq 0 \neq n_0 $, and additionally $n_0 = m_0$
 and  $\alpha_{10} = 0$ as well.
This corresponds to a physical situation where a system with total angular momentum zero is decomposed or decays into two subsystems which move in opposite
 directions, one 
with angular momentum $m_0$ and the other with $m_1= -m_0$. Example is a neutral particle with spin zero (e.g., a neutral pion or a Higgs particle) which - 
in its rest frame - decays into 2 photons
with opposite spins one.

 With  the simplifications mentioned we get for the expression \eqref{eq:31}:  {\allowdisplaybreaks
\begin{align}
  \label{eq:34}
 & (2\pi)^2V_{\psi_{2\!-\!qb}}(\tl{\theta},\tl{p}) \\ & = b_{00}^2\,\sinc\pi(p_1-m_0)\sinc\pi(p_2-m_0)\nonumber \\ & +b_{10}^2\sinc\pi(p_1+m_0)\sinc\pi(p_2-m_0) \nonumber\\
&+b_{01}^2\sinc\pi(p_1-m_0)\sinc\pi(p_2+m_0)\nonumber \\ & +b_{11}^2\sinc\pi(p_1+m_0)\sinc\pi(p_2+m_0) \nonumber \\
 &+ 2\,b_{00}b_{10}\cos(2m_0\theta_1) 
 \sinc\pi p_1\sinc\pi(p_2-m_0) \nonumber \\
&+ 2\,b_{00}b_{01}\cos(2m_0\theta_2+ \alpha_{01}) 
 \sinc\pi(p_1-m_0)\sinc\pi p_2 \nonumber \\
&+ 2\,b_{00}b_{11}\cos[2m_0(\theta_1+ \theta_2)+\alpha_{11}] 
 \sinc\pi p_1\sinc\pi p_2 \nonumber\\
&+ 2\,b_{01}b_{10}\cos[2m_0(\theta_1-\theta_2)- \alpha_{01}] 
 \sinc\pi p_1\sinc\pi p_2 \nonumber \\
&+ 2\,b_{01}b_{11}\cos(2m_0\theta_1+ \alpha_{11}-\alpha_{01})
 \sinc\pi p_1\sinc\pi(p_2+m_0) \nonumber \\
&+ 2\,b_{10}b_{11}\cos(2m_0\theta_2+ \alpha_{11})
 \sinc\pi(p_1+m_0)\sinc\pi p_2 \nonumber.
\end{align}}
This expression will be useful below (see example 6).
\subsubsection{Wigner functions for non-factorizable 2-qubits: EPR/Bell states}
 The non-factorizable EPR/Bell states \eqref{eq:4} or \eqref{eq:32}  are special cases of the states \eqref{eq:28} with either $b_{00}b_{11} \neq 0,\,b_{10}=0=b_{01}$ or
 $b_{10}b_{01} \neq 0,\,b_{00}=0=b_{11}$, i.e.
 \begin{equation}
   \label{eq:63}
   \psi^{(\alpha_{11}, \beta)}_{00,11}(\tl{\vp}) =   \cos\beta e_{m_0n_0}(\tl{\vp})+ \sin\beta\, e^{i\alpha_{11}}\,e_{m_1n_1}(\tl{\vp})
 \end{equation}
 and
 \begin{equation}
   \label{eq:64}
   \psi^{( \alpha_{01}, \gamma)}_{10,01}(\tl{\vp}) =   \cos\gamma\, \ e_{m_1n_0}(\tl{\vp})+ \sin\gamma\, e^{i\alpha_{01}}\,e_{m_0n_1}(\tl{\vp}).
 \end{equation} 
 According to Eq.\ \eqref{eq:31} the Wigner function of the state \eqref{eq:63} is given by
\begin{align}
  \label{eq:52}
   & V_{00,11}^{(\alpha_{11},\beta)}(\tl{\theta},\tl{p}) \\ & = \frac{1}{(2\pi)^2} \{\cos^2\beta \sinc\pi(p_1-m_0)\sinc\pi(p_2-n_0)  \nonumber
 \\ & + \sin^2\beta \sinc\pi(p_1-m_1)\sinc\pi(p_2-n_1)
\nonumber \\ & +
\sin 2\beta \cos[(m_1-m_0)\theta_1+ (n_1-n_0)\theta_2 +\alpha_{11}] \nonumber \\ & \times \sinc\pi[p_1-(m_0+m_1)/2]\sinc\pi[p_2-(n_0+n_1)/2]\}. \nonumber
\end{align}
with a corresponding expression for the Wigner function of the state \eqref{eq:64}:
\begin{align}
  \label{eq:65}
   & V_{10,01}^{(\alpha_{01},\gamma)}(\tl{\theta},\tl{p}) \\ & = \frac{1}{(2\pi)^2} \{\cos^2\gamma \sinc\pi(p_1-m_0)\sinc\pi(p_2-n_0)  \nonumber
 \\ & + \sin^2\gamma \sinc\pi(p_1-m_1)\sinc\pi(p_2-n_1)
\nonumber \\ & +
\sin 2\gamma \cos[(m_1-m_0)\theta_1- (n_1-n_0)\theta_2 -\alpha_{01}] \nonumber \\ & \times \sinc\pi[p_1-(m_0+m_1)/2]\sinc\pi[p_2-(n_0+n_1)/2]\}. \nonumber
\end{align}
The Wigner functions \eqref{eq:52} and \eqref{eq:65} have the following essential properties:

1. They are not factorizable into two factors each of which depends on $\theta_1$ or $\theta_2$ only.

2. The $\theta_1$ - and $\theta_2$ - dependences of the expression \eqref{eq:52} are solely given by the argument
\begin{equation}
  \label{eq:66}
  \vt_+(\tl{\theta}) = (m_1-m_0)\theta_1+ (n_1-n_0)\theta_2 +\alpha_{11}
\end{equation}
of the {\it non-classical $\cos$-interference term} and that of the function \eqref{eq:65} correspondingly by

\begin{equation}
  \label{eq:67}
  \vt_{-}(\tl{\theta})= (m_1-m_0)\theta_1- (n_1-n_0)\theta_2 -\alpha_{01};
\end{equation}
with the inversions
\begin{align}\label{eq:68}
  \theta_1 &= \frac{1}{2(m_1-m_0)}[\vt_+ + \vt_- + \alpha_{01} - \alpha_{11}], \\
   \theta_2 &= \frac{1}{2(n_1-n_0)}[\vt_+ - \vt_- + \alpha_{01} + \alpha_{11}]. \label{eq:69}
\end{align}
Note that here $m_1-m_0 \neq 0 \neq n_1 - n_0$.

3. As the function \eqref{eq:52} does not depend on $\vt_-$ it is instructive to consider the configuration space curve
\begin{align}
  \label{eq:49}
  \mathcal{C}_{+} =& \{(\theta_{1}=\theta_1(\vt_+), \theta_2 =\theta_2(\vt_+)\,) \in \mathbb{S}^1 \times \mathbb{S}^1, \\& \vt_+\in \mathbb{R};\,\,                                                                                                   \vt_-,\, \alpha_{01},\,\alpha_{11} = \text{const.}\}.  \nonumber
\end{align}

Thus, the domain of the curve \eqref{eq:49} is a spiral around a torus with  ``slopes'' $\partial \theta_1/\partial \vt_+ = 1/[2(m_1-m_0)]$                         and $\partial \theta_2/\partial \vt_+ = 1/[2(n_1-n_0)]$.

 {\em The above properties 1.-3. show that the non-factorizà\-bility of EPR/Bell states is associated with the non-factorizability of the corresponding Wigner function.

    The Wigner functions \eqref{eq:52} and \eqref{eq:65}  do depend on the configuration space variables
 $\theta_1$ and $\theta_2$  only through the  non-classical interference terms $\cos \vt_+(\theta_1, \theta_2)$ or $\cos \vt_-(\theta_1, \theta_2)$, in such a way that their domain is a curve which spirals around a torus. Thus, characteristic features of A-OAM Wigner functions for EPR/Bell states are related to the
 non-trivial topological properties of a torus, the configuration subspace of the total phase space.} 

\subsubsection{Marginal distributions}
The marginal (separate) probability distribuions for the angles $\tl{\theta}$  and the momenta $\tl{p}$, respectively, can - like in subsection II.B. - be obtained by integrating, e.g., the Wigner function \eqref{eq:52} over the  complementary canonical variables:
\begin{align}
  \label{eq:50}
  & \int_{-\infty}^{\infty}dp_1dp_2 V_{00,11}^{(\alpha_{11}, \beta)}(\tl{\theta},\tl{p})\\ & = \frac{1}{(2\pi)^2}[1+\sin 2\beta \cos[\vt_+(\tl{\theta})],
\nonumber  \end{align}
where the relation \eqref{eq:14} has been used. The expression \eqref{eq:50} equals $1/(2\pi)^2| \psi^{(\alpha_{11}, \beta)}_{00,11}(\tl{\vp})|^2$ with $\psi$ of
Eq.\ \eqref{eq:63} and $\tl{\vp}$ replaced by $\tl{\theta}$. The factor $1/(2\pi)^2$ corresponds to the same factor in the integral \eqref{eq:24}.

Integrating the function \eqref{eq:52} over $\tl{\theta}$ yields the Whittaker cardinal function\cite{ka2}
\begin{align}
  \label{eq:70}
  & \int_{-\pi}^{\pi}d\theta_1d\theta_2 V_{00,11}^{(\alpha_{11}, \beta)}(\tl{\theta},\tl{p})\\ & = \cos^2\beta [\sinc\pi(p_1-m_0)\sinc\pi(p_2+n_0)] \nonumber \\
&+ \sin^2\beta [\sinc\pi(p_1-m_1)\sinc\pi(p_2-n_1)]\nonumber\\ & = \omega_{00,11}(\tl{p}). \nonumber 
\end{align}
Again, using the orthonormality relations \eqref{eq:16} the probabilities $|c_{00}|^2=\cos^2\beta$ and $|c_{11}|^2 =\sin^2\beta$ for finding the OAM pairs $(m_0,m_0)$ or $(m_1,n_1)$,
respectively,
can be obtained from the Whittaker cardinal function $\omega_{00,11}(\tl{p})$ of Eq.\ \eqref{eq:70}.

The Wigner function \eqref{eq:65} can be treated accordingly. 

\subsubsection{Wigner functions of the EPR/Bell states basis}  

\begin{figure}[h]\includegraphics[width=\columnwidth]{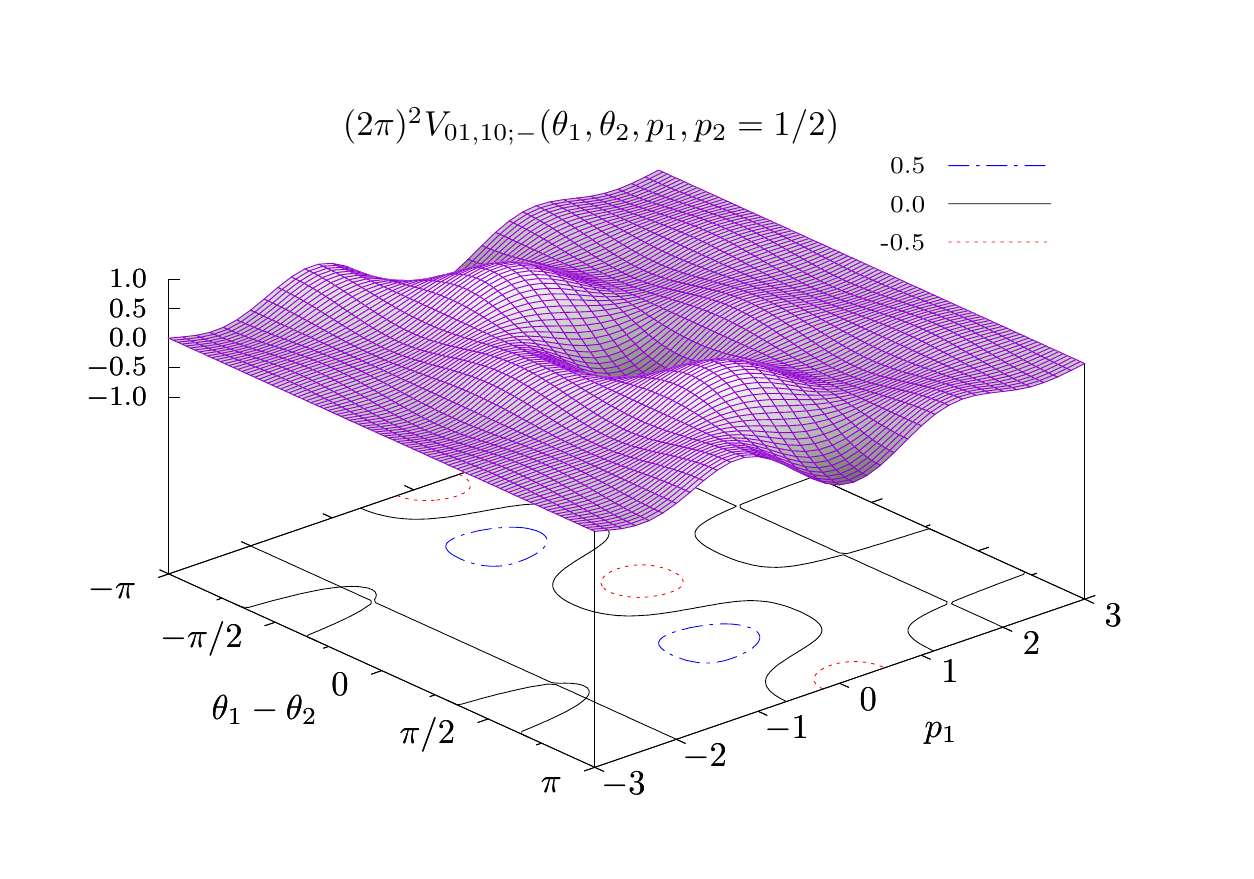}
\caption{\label{fig:fig3c} Wigner function $(2\pi)^2 V_{01,10;-}(\theta_1, \theta_2,p_1, p_2 = 1/2) = \{-\frac{1}{3}\sinc\pi(p_1 -1) + \sinc\pi(p_1 + 1)
-2\cos[2(\theta_1-\theta_2)]\sinc\pi p_1\}/\pi$ of the EPR/Bell state $\psi_{01,10;-}(\tl{\vp})$ from Eq.\ \eqref{eq:37}.} \end{figure}

According to the correspondences \eqref{eq:33} we get  for the basic EPR/Bell states \eqref{eq:4} and \eqref{eq:32} the functions (with $n_0 = m_0 \in \mathbb{Z} -
\{0\},\,n_1 = m_1= -m_0$)
\begin{align}
  \label{eq:41}
  &\psi_{00,11;\pm}(\tl{\vp})\\ & = \frac{1}{\sqrt{2}}[e_{m_0}(\vp_1)e_{m_0}(\vp_2)\pm e_{-m_0}(\vp_1)e_{-m_0}(\vp_2)], \nonumber \\
&~~~~~ L\, \psi_{00,11;\pm}(\tl{\vp}) = 2(m_0 \mp m_0) \psi_{00,11;\pm}(\tl{\vp}); \\ \label{eq:42}
  &\psi_{01,10;\pm}(\tl{\vp})\\ & = \frac{1}{\sqrt{2}}[e_{m_0}(\vp_1)e_{-m_0}(\vp_2)\pm e_{-m_0}(\vp_1)e_{m_0}(\vp_2)], \nonumber \\
&~~~~~L\, \psi_{01,10;\pm}(\tl{\vp}) = 0. 
\end{align}
Taking $\beta = \pi/4,\, \alpha_{11} =0,\,\pi$ in Eq.\ \eqref{eq:52} and $\gamma = \pi/4,\, \alpha_{01} =0,\, \pi$ in Eq.\ \eqref{eq:65} gives the associated Wigner functions (see also the expression \eqref{eq:34})
\begin{align}
  \label{eq:36}
    (2\pi)^2& V_{00,11;\pm}(\tl{\theta},\tl{p})\\ & = \frac{1}{2}[\sinc\pi(p_1-m_0)\sinc\pi(p_2-m_0)] \nonumber \\
 &+ \frac{1}{2}[\sinc\pi(p_1+m_0)\sinc\pi(p_2+m_0)]\nonumber \\
 & \pm \cos[2 m_0(\theta_1+ \theta_2)]\sinc\pi p_1\sinc\pi p_2; \nonumber \\
    (2\pi)^2& V_{01,10;\pm}(\tl{\theta},\tl{p}) \label{eq:37} \\ & = \frac{1}{2}[\sinc\pi(p_1-m_0)\sinc\pi(p_2+m_0)] \nonumber \\
 &+ \frac{1}{2}[\sinc\pi(p_1+m_0)\sinc\pi(p_2-m_0)]\nonumber \\
 & \pm \cos[2m_0(\theta_1- \theta_2)]\sinc\pi p_1\sinc\pi p_2. \nonumber
\end{align}
Let us look at special cases of the expression \eqref{eq:37}, with the minus sign in Eq.\ \eqref{eq:42} and  the related one in Eq.\ \eqref{eq:37} as well:

An example of such functions \eqref{eq:37} with $m_0 =1$ and $p_2 = 1/2$ is shown in Fig.\ \ref{fig:fig3c}.

\begin{figure}[h]\includegraphics[width=\columnwidth]{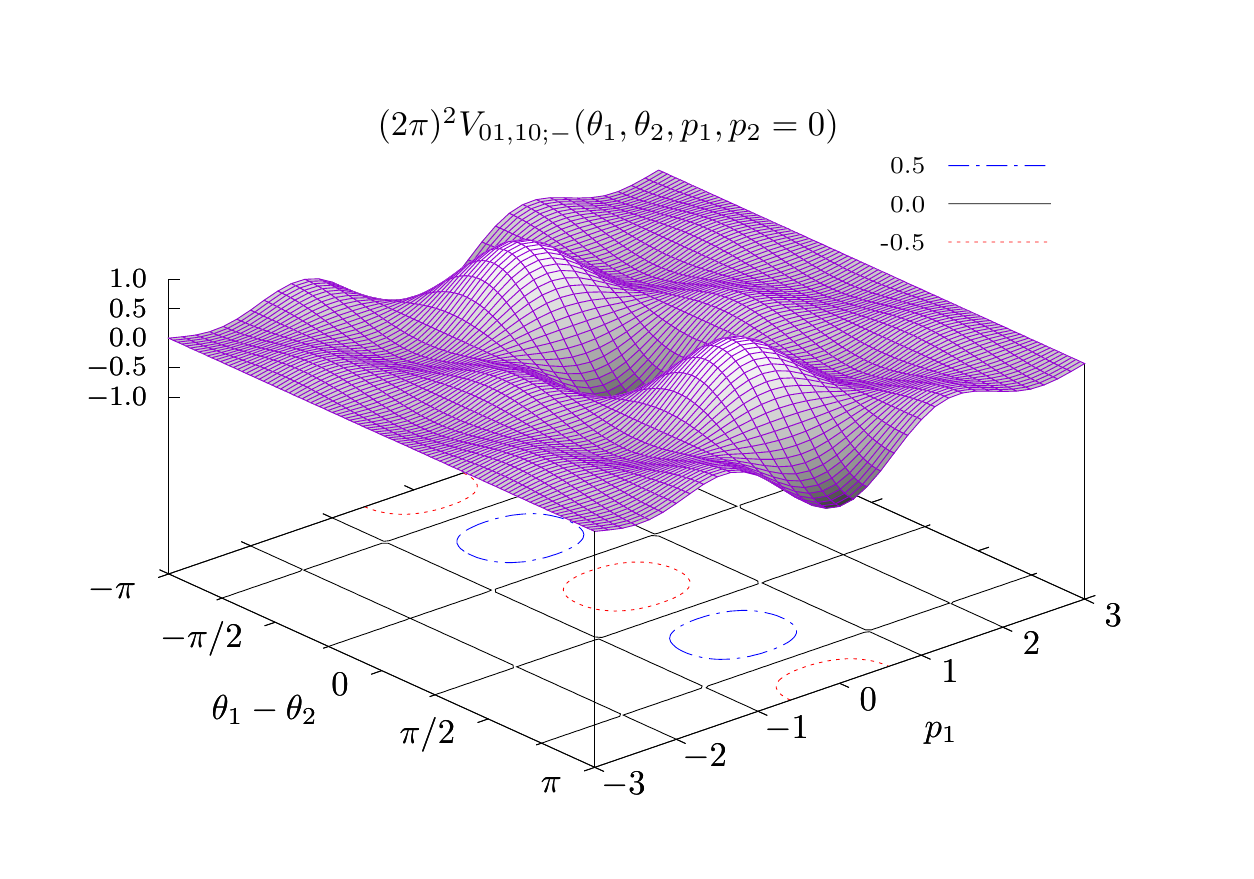}
\caption{\label{fig:fig4c} Wigner function $(2\pi)^2 V_{01,10;-}(\theta_1, \theta_2,p_1, p_2 = 0) = 
-\cos[2(\theta_1-\theta_2)]\sinc\pi p_1$, according to Eq.\ \eqref{eq:40} with $m_0 =1$, of the EPR/Bell state $\psi_{01,10;-}(\tl{\vp})$ from Eq.\ \eqref{eq:42}.}
\end{figure}

Other examples are
\begin{equation}
  \label{eq:48}
  V_{01,10;-}(\tl{\theta},p_1,p_2) = 0 \text{ for } p_1,p_2 \in \{\mathbb{Z} - \{0\}\\\},
\end{equation}
and
\begin{align}
  \label{eq:40}
 (2\pi)^2 V_{01,10;-}&(\tl{\theta},p_1,p_2=0)\\ & = -\cos[2m_0(\theta_1-\theta_2)]\sinc\pi p_1, \nonumber
\end{align}
 with the corresponding relation for $p_1 =0$. 
A graphical illustration of the function \eqref{eq:40} with $m_0 =1$ is given in Fig.\ \ref{fig:fig4c}.

Furthermore,
\begin{equation}
  \label{eq:46}
  V_{01,10;-}(\theta_1, \theta_2, p_1=0, p_2=0) = -\frac{1}{4\pi^2}\cos[2m_0(\theta_1-\theta_2)],
\end{equation}
showing explicitely that the Wigner function is negative on certain subsets of the phase space.

\begin{acknowledgments} I very much thank the DESY Theory Group for its sustained and very kind hospitality
after my retirement from the Institute for Theoretical Physics of the RWTH Aachen. I am grateful to Hartmann R\"{o}mer for
a fruitful  discussion and to David
Kastrup for providing the figures. I thank Jakub Rembieli\'{n}ski and Krzystof Kowalski for the invitation to their Theoretical Physics Institute at the University of {\L}\'odz (May 2017) and for many stimulating discussions with them. Equally I thank Gerd Leuchs and Luis Sanchez-Soto for their invitation to the MPI for the Science of Light, Erlangen, and for many fruitful discussions (February 2018). Finally I am obliged to family and friends for help and encouragements during a longer period of illnesses. \end{acknowledgments}

\begin{appendix}
  \section{Fractional OAM with $\dl \neq 0$}
  \subsection{Generalities}
 In the main text above we have discussed A-OAM Wigner functions in the Hilbert space framework for the special case $\dl =0$ of the  more general frameworks as described by the Eqs.\ \eqref{eq:6}-\eqref{eq:2}. The mathematical and physical meanings of the  ``fractional'' case $\dl \in (0,1)$ were discussed in Ref.\cite{ka}. In the present appendix the generalization of the main  results for Wigner functions with $\dl =0$ to the more general case $\dl \neq 0$ will be sketched (for the same topic see also Subsection V.A. of Ref.\ \cite{ka2}):

 The essential recipe for doing so are the replacements
 \begin{equation}
   \label{eq:53}
   m_0,\, m_1 \rightarrow m_0 + \dl,\, m_1 + \dl 
 \end{equation}
 in Section II and
 \begin{equation}
   \label{eq:54}
    m_0, m_1 \rightarrow m_0 + \dl_1, m_1 + \dl_1, \,\, n_0, n_1 \rightarrow n_0 + \dl_2, n_1 + \dl_2  
  \end{equation}
  in Section III, where the two Hilbert spaces of the tensor product may be different.

  Instead of the state \eqref{eq:7} we now have
  \begin{align}
    \label{eq:71}
   \chi_{m_0,m_1}^{[\dl](\alpha, \beta)}(\vp) =& \cos \beta\, e_{m_0,\dl}(\vp) + \sin \beta\, e^{i \alpha}\,e_{m_1,\dl}(\vp),\\ & m_0,m_1 \in \mathbb{Z},\; m_1 \neq m_0. \nonumber  
  \end{align}
  Its absolute value squared is the same as in Eq.\ \eqref{eq:47}.

  In Eqs.\ \eqref{eq:27} and \eqref{eq:60} the $e_{m_j}$ have to be replaced by $e_{m_j,\dl}$.

  The expressions \eqref{eq:9} and \eqref{eq:22} become now
  \begin{align}
  \label{eq:72}
  V_{\psi}^{[\dl]}(\theta,p) & = \frac{1}{2\pi}\int_{-\pi}^{\pi}\frac{d\vt}{2\pi}e^{-i (p/\hbar)\vt}\psi^{[\dl]\ast}(\theta\!-\!\frac{\vt}{2})\psi^{[\dl]}(\theta\! +\!\frac{\vt}{2}) \nonumber \\ &=                                                                                                                                                                            \sum_{m,n \in \mathbb{Z}} c_m^{\ast}V_{mn}^{[\dl]}(\theta,p)c_n\,, \\ V_{mn}^{[\dl]}(\theta,p) & = \frac{1}{2\pi}e^{i(n-m)\theta}\int_{-\pi}^{\pi}\frac{d\vt}{2\pi}
                                                                                                                                                                                                                                                              e^{i[(m+n+ 2\dl)/2-p/\hbar]\vt} 
\label{eq:73} \\ 
                             & = \frac{1}{2\pi}e^{i(n-m)\theta}\sinc \pi[p/\hbar - (m+n+ 2\dl)/2], \nonumber
  \end{align}
  which implies for the Wigner function \eqref{eq:10} the generalization
\begin{align}
  \label{eq:74}
 & 2\pi\,V_{m_0,m_1}^{[\dl](\alpha, \beta)}(\theta,p) \\ &=  \cos^2\beta \sinc\pi(p-m_0-\dl) + \sin^2\beta \sinc\pi(p-m_1-\dl)\nonumber \\
&+ \sin 2\beta \cos[(m_0-m_1)\theta - \alpha] \nonumber \\ & \times \sinc\pi[p-(m_0 + m_1 + 2\dl)/2].\nonumber 
\end{align}
Note that the $\theta$-dependent $\cos$-interference term remains independent of $\dl$, whereas the different $\sinc$-functions have their maximum $1$ now at
$p = m_0 + \dl, \; p= m_1 + \dl$ and $p= (m_0 + m_1 + 2\dl)/2$.

The additional resulting changes for the formulae in Subsections II.A. and II.B. due to the replacements \eqref{eq:53} are obvious. The relation \eqref{eq:16} also
remains valid under those substitutions.

For the mixed states of Subsection II.C. we obtain instead of Eq.\ \eqref{eq:55}
\begin{align}
  \label{eq:75}
  V_{\rho^{[m_0,m_1]}}^{[\dl]}(\theta,p) =& \tr\{\rho^{[m_0,m_1]}\cdot V^{[\dl;m_o,m_1]}(\theta, p)\}, \\ & V^{[\dl; m_0,m_1]}(\theta, p)= (V_{m_j m_k}^{[\dl]}(\theta, p)), \nonumber
\end{align}
with the matrix elements $V_{m_j m_k}^{[\dl]}(\theta,p),\, j,k = 0,1 $ from Eq.\ \eqref{eq:73}. The main result \eqref{eq:58} is replaced by
\begin{align}
  \label{eq:76}
 & 2\pi \tr[\rho^{[m_0, m_1]}\cdot V^{[\dl; m_0, m_1]}(\theta, p)]\\& =  \frac{1+ a_3}{2} \sinc \pi (p-m_0-\dl) + \frac{1-a_3}{2}\sinc \pi(p-m_1-\dl) \nonumber \\ & 
+ \{a_1\cos[(m_0-m_1)\theta] +
a_2\sin[(m_0 -m_1)\theta]\}  \nonumber \\ & \times \sinc \pi [p - (m_0 + m_1 +2\dl)/2]. \nonumber
\end{align}
The generalizations of the formulae in Section III are slightly more complicated because the Hilbert space factors of Eq.\ \eqref{eq:39} may have different $\dl_1$ and $\dl_2$:
\begin{equation}
  \label{eq:77}
  L^2(\mathbb{S}^1\times \mathbb{S}^1, d\vp_1d\vp_2/(2\pi)^2;\dl_1,\dl_2),
\end{equation}
 with basis
\begin{align}
  \label{eq:78}
  e_{mn;\dl_1,\dl_2}(\tl{\vp}) =& e_{m,\dl_1}(\vp_1)\,e_{n,\dl_2}(\vp_2)\\ =& e^{i(m+\dl_1)\vp_1 + i(n+\dl_2)\vp_2},\, m,n \in \mathbb{Z}, \nonumber
\end{align}
scalar product
\begin{align}
  \label{eq:79}
  (\psi_2^{[\dl_1,\dl_2]},\psi_1^{[\dl_1,\dl_2]}) =& \int_{-\pi}^{\pi}\frac{d^2\tl{\vp}}{(2\pi)^2}\psi^{[\dl_1,\dl_2]\ast}_2(\tl{\vp})\psi_1^{[\dl_1,\dl_2]}(\tl{\vp}),\\
&(e_{km;\dl_1,\sl_2},e_{ln;\dl_1,\dl_2}) = \delta_{kl}\,\delta_{mn}, \nonumber
\end{align}
and expansions
\begin{align}
  \label{eq:80}
  \psi^{[\dl_1,\dl_2]}(\tl{\vp}) =& \sum_{m,n \in \mathbb{Z}}c_{mn}\,e_{mn;\dl_1,\dl_2}(\tl{\vp}),\\ & c_{mn}  = (e_{mn;\dl_1,\dl_2},\psi^{[\dl_1,\dl_2]}). \nonumber
\end{align}
The functions \eqref{eq:78} are eigenfunctions of the total OAM operator:
\begin{equation}
  \label{eq:81}
  L = \frac{1}{i}\partial_{\vp_1} +\frac{1}{i}\partial_{\vp_2},\;\; Le_{mn;\dl_1,\dl_2} = (m+n+\dl_1 +\dl_2)e_{mn;\dl_1,\dl_2}. 
\end{equation}
Instead of the correspondences \eqref{eq:33} we now have
\begin{align}
  \label{eq:82}
 &e_{m_0 n_0;\dl_1,\dl_2}\leftrightarrow |00\rangle,\;  e_{m_1 n_1;\dl_1,\dl_2}\leftrightarrow |11\rangle,\\
&  e_{m_0 n_1;\dl_1,\dl_2}\leftrightarrow |01\rangle,\;  e_{m_1 n_0;\dl_1,\dl_2}\leftrightarrow |10\rangle. \nonumber 
\end{align}
In the Wigner function \eqref{eq:26} the wave function $\psi$ has to be replaced by $\psi^{[\dl_1,\dl_2]}$ from Eq.\ \eqref{eq:80}:
\begin{align}
  \label{eq:83}
 & (2\pi)^2V_{\psi}^{[\dl_1,\dl_2]}(\tl{\theta}, \tl{p}) =  \\ & \int_{-\pi}^{\pi}\frac{d^2\tl{\vt}}{(2\pi)^2}
e^{\!-\!i(p_1\vt_1 + p_2\vt_2)}\,\psi^{[\dl_1,\dl_2]\ast}(\tl{\theta}\!-\!\tl{\vt}/2)
 \psi^{[\dl_1,\dl_2]}(\tl{\theta}\!+\!\tl{\vt}/2). \nonumber
\end{align}
The general 2-qubit elements \eqref{eq:28} are replaced by
\begin{align}
  \label{eq:84}
&  \psi_{2\!-\!qb}^{[\dl_1,\dl_2]}(\tl{\vp})  =\\ &c_{00}\,e_{m_0 n_0;\dl_1,\dl_2}(\tl{\vp}) +c_{10}\,e_{m_1 n_0;\dl_1,\dl_2}(\tl{\vp})\nonumber \\& + c_{01}\,e_{m_0 n_1;\dl_1,\dl_2}(\tl{\vp}) +
    c_{11}\,e_{m_1n_1;\dl_1,\dl_2}(\tl{\vp}), \nonumber \\
&|c_{00}|^2 + |c_{01}|^2 + |c_{10}|^2 + |c_{11}|^2 = 1. \nonumber
\end{align}
Inserting this state into the  expression \eqref{eq:83} gives the obvious generalization of the 2-qubit Wigner function \eqref{eq:31}.

It is remarkable that the $\theta$-dependent $\cos$-factors in all interference terms of the general expression \eqref{eq:31} remain unchanged under the substitutions \eqref{eq:54}. See also the corresponding invariances of the Wigner functions \eqref{eq:52} and \eqref{eq:65} and those of the arguments \eqref{eq:66} and \eqref{eq:67} of their
$\cos$-factors.

For, e.g., the Wigner function \eqref{eq:52} we have explicitly
\begin{align}
  \label{eq:85}
   & (2\pi)^2 V_{00,11}^{[\dl_1,\dl_2](\alpha_{11},\beta)}(\tl{\theta},\tl{p}) \\ & =  \cos^2\beta \sinc\pi(p_1-m_0-\dl_1)\sinc\pi(p_2-n_0-\dl_2)  \nonumber
 \\ & + \sin^2\beta \sinc\pi(p_1-m_1-\dl_1)\sinc\pi(p_2-n_1-\dl_2)
\nonumber \\ & +
\sin 2\beta \cos[(m_1-m_0)\theta_1+ (n_1-n_0)\theta_2 +\alpha_{11}] \nonumber \\ & \times \sinc\pi[p_1-(m_0+m_1+2\dl_1)/2]\nonumber \\ & \times \sinc\pi[p_2-(n_0+n_1+2\dl_2)/2]\}. \nonumber
\end{align}
Thus, the $\theta$-dependence of the  Wigner functions as defined on classical phase space is not affected by the substitutions \eqref{eq:53} and \eqref{eq:54}. The changes of the momenta $p$ and $\tl{p}$ are intuitively obvious.
\vspace{0.5cm}
\subsection{An important application}
An essential case of a $\dl \neq 0$ Hilbert space was already discussed in the context of example 3 in Subsection IV.C. of Ref.\ \cite{ka2}:
Suppose $\psi_b(\vp)$ is a function of $\vp$ with the property
\begin{equation}
  \label{eq:86}
  \psi_b(\vp + 2\pi) = e^{i 2\pi b \vp}\,\psi_b(\vp),\, b \in \mathbb{R}.
\end{equation}
The real number $b$ can be decomposed uniquely into an integer plus a $\dl$ of Eq.\ \eqref{eq:6}:
\begin{equation}
  \label{eq:87}
  b = n_b + \dl,\; \dl \in [0,1),\; n_b \in \mathbb{Z}.
\end{equation}
If the integral
\begin{equation}
  \label{eq:88}
   \int_{-\pi}^{\pi}\frac{d\vp}{2\pi}\psi^{\ast}_b(\vp)\psi_b(\vp)
\end{equation} exists then $\psi_b(\vp)$ can be considered as an element of $L^2(\mathbb{S}^1, d\vp/2\pi;\dl)$. $\psi_b(\vp)$ is a kind of Bloch wave\cite{ka}.
\end{appendix}
\bibliography{wignerfunction}
\end{document}